\documentclass[a4paper,12pt,oneside]{article}
\usepackage{graphicx}

\begin{document}

\title{The model of dynamo with small number of modes and magnetic activity
of T Tauri stars}

\author{D.D.Sokoloff$^1$, S.N.Nefedov$^1$, A.A.Ermash$^2$, S.A.Lamzin$^2$ }

\date{ $^1$ Department of Physics, Moscow State University, Moscow, Russia;\\
$^2$ Sternberg Astronomical Institute, Moscow, Russia 
\footnote {Accepted by Astron. Letters, E-mail: sokoloff@dds.srcc.msu.su,
lamzin@sai.msu.ru}
\\
}

\maketitle

\bigskip
\section*{Abstract}{
   The model that describes operation of dynamo in fully convective stars is
presented. It is based on representation of stellar magnetic field as a
superposition of finite number of poloidal and toroidal free damping modes.
In the frame of adopted low of stellar differential rotation we estimated
minimal value of dynamo number D, starting from which generation of cyclic
magnetic field in stars without radiative core is possible. We also derived
expression for period of the cycle. It was found that dynamo cycles of fully
convective stars and stars with thin convective envelopes differ in a
qualitative way: 1) distribution of spots over latitude during the cycle is
different in these stars; 2) the model predicts that spot formation in fully
convective stars should be strongly suppressed at some phases of the cycle.

  We have analyzed historical lightcurve of WTTS star V410 Tau and found
that long term activity of the star is not periodic process. Rather one can
speak about quasi cyclic activity with characteristic time of $\sim 4$ yr
and chaotic component over imposed. We concluded also that redistribution of
cool spots over longitude is the reason of long term variations of V410 Tau
brightness. It means that one can not compare directly results of
photometric observations with predictions of our axially symmetric (for
simplicity) model which allows to investigate time evolution of spot's
distribution over latitude. We then discuss what kind of observations and in
which way could be used to check predictions of the dynamo theory.
}

\bigskip
\section*{Introduction}{}
\bigskip

  Origin of large-scale magnetic field is mainly connected with action of
dynamo based on differential rotation and so-called $\alpha$-effect which
can be explained as follows (Parker, 1955). Differential rotation produces
toroidal magnetic field component from poloidal one while $\alpha$-effect
produces poloidal magnetic field from toroidal one to close the chain of
self-excitation. As a result, a dynamo wave of magnetic field as well as
corresponding surface activity wave are excited and give solar 11-year
(better to say 22-year) cycle.

  From the first sight, it looks attractive to present the above poloidal
field as magnetic field of magnetic dipole prolongated into the stellar
interior in the way which gives inside a rigid rotating star without
$\alpha$-effect (i.e. without magnetic field generation) a magnetic
configuration with decays in time as slow as possible. Such dipole field is
known in electrodynamics as basic free decay mode. Similarly, it looks
attractive to present the above toroidal field naively as a toroidal field
which decays in time as slow as possible in a star without sources of
generation. Both fields has as few zeros as possible for poloidal and
toroidal fields correspondingly in its latitudinal profiles. In fact however
dynamo excites much more complicated spatial magnetic field configurations
which latitudinal profile has more zeros rather above mentioned simple ones.
These configurations can be presented as linear combinations of
eigenfunctions of Maxwell equations for a star without any source of
generation, i.e. differential rotation and $\alpha$-effect. These 
eigenfunctions are known as free decay modes.

  The above generation mechanism is applicable to any star with a pronounced
convec-tive shell. On the other hand, observations as well as computer
simulations give hints that dynamo in a fully convective star has some
specific features. We are going to demonstrate that these features are
related to the fact that magnetic configurations generated by dynamo in
fully convective star can be represented as a combination of free decay
modes different from that one for a magnetic field dynamo generated in a
thin shell. In particular, the number of zeros in latitudinal profiles for
both sets of free decay modes are specific for each case.

   Low mass $(M \le 1.5 M_\odot)$ young stars $(t\le 10$ Myr) are of special
interest in this connection because they are fully convective or have
external convective zone that much larger than solar and their
chromospheric-coronal activity is very high. We'll discuss below only young
stars whose activity results from interaction of stellar magnetic field with
convective zone's matter solely, i.e. we'll speak about so called weak line
T Tauri stars (WTTSs). Near two thousands WTTSs are known at the moment.
\footnote{ Probably the nature of activity of brown dwarfs
$(M\le 0.08 M_\odot)$ without accretion disks is the same but we shall not
discuss these objects due to deficit of observational data.}

  Activity of WTTSs is manifested as emission in Ca\,II H, K and H$_\alpha$
lines, thermal X-ray and non-thermal radio emission, as well as flares in
optical and/or X-ray bands (Bertout, 1989). In average flares occur 100
times more frequently in WTTS than in the Sun and their average energy is
100 times larger than average solar flares, such as flares with total energy
liberation up to $3\cdot 10^{37}$ erg were observed (Fernandez et al., 2003;
Feigelson et al., 2005). Periodical variations of brightness resulted from
existence of cold $(T<T_{\rm ef})$ spots on WTTS's surface is a typical
feature of these objects.  Spots can cover up to 30-40 \% of stellar
surface, save its shape during hundreds rotational periods and can locate
both in equatorial and polar regions (Strassmeier et al., 1994; Joncour et
al., 1994a,b; Hatzes, 1995; Rice \& Strassmeier, 1996).

   Angular velocity of WTTSs in average an order of magnitude larger than
solar (Herbst et al., 2005) and only in one case differential rotation was
observed (Herbst et al., 2006). Surface magnetic field strength of WTTSs is
$\sim 1-3$ kG, but it is not still clear if these stars have global magnetic
field or observed field is a superposition of small scale fields of numerous
active regions -- see Johns-Krull et al. (2004) and references therein. Week
dependence of X-ray to bolometric luminocities ratio on rotation period is a
typical feature of WTTSs probably because active regions cover almost all
surface of these stars (Feigelson et al., 2005). Thus activity of WTTSs
differs from solar activity not only in a quantitative but also in a
qualitative way, what looks reasonable to connect with specific character of
magnetic field generation process in these stars.

  Meanwhile it is not still clear if WTTS's magnetic field is generated by
dynamo or it is a relic field of protostellar cloud enhanced during
contraction -- see e.g. Dudorov (1995). Detection of activity cycles in
young stars would be a serious argument in favor of dynamo operation,
especially if the theory could reproduce parameters of these cycles.

  One can try to envisage specific features of dynamo action in fully
convective stars via direct numerical simulations -- see e.g. Dobler et al.
(2006). But being attractive by itself, this approach has limitations: one
have to adopt many relevant parameters to describe stellar hydrodynamics
while many of them are still not known from observations. To be specific,
note that we are interested here in radial profiles of the angular
velocities in the whole extent of the convective shell while observations
give at most the longitudinal dependence at the stellar surface. It is why
we are interested in simple qualitative models of generation confronted with
corresponding models for thin convective shells as well as with
observations. This is the aim of the paper.

\section*{Low mode approximation for stellar dynamo}

  In the frame of the proposed simplest model we consider growing stellar
magnetic field as a combination of dipole and a few following multipoles,
considered as free decay modes. The number of these modes should be
subsequently enlarged until one obtains the solution which grows in time.

  Free decay modes for fully convective star are presented in the
fundamental Moffatt's (1980) book, where they are enumerated according to
the time decay rate as follows. Poloidal singlet $P_1$ (which is referred as
dipole) has the lowest decay rate, then comes the doublet consisting of the
second poloidal mode and the first toroidal mode $P_2+T_1$, then comes the
singlet $P_3$, the doublet $P_4+T_2$ etc. For simplicity we considered
axisymmetric modes only while it is used to suppose that magnetic axis of T
Tauri stars can be substantially inclined to the rotation axis (e.g. Bouvier
et al, 2007).

 Let fix a few first free decay modes and look for a magnetic field in a
form of a linear combination of the modes with coefficients to be obtained
from dynamo equations. Substituting the Ansatz in the dynamo equations
(Krause and R\"adler, 1980) we arrive to an eigenvalue problem for the
vector $\bf C$ composed from the coefficients:
\begin{equation}
 (\gamma E - M) C = \hat W C.
  \label{dyn}
\end{equation}
Here $\gamma$ is a complex growth rate, imaginary part of which determines
the length of activity cycle; $E$ is the unit matrix; $M$ is a diagonal
matrix, non-zero elements of which $M_{ii}$ are decay rates $\gamma_i$ of
considered modes; $C$ is a column, elements of which are amplitudes of
modes of interest. Elements of $\hat W$ matrix are scalar products:
$$
W_{ij} = \int {\bf B}^*_i \hat L {\bf B}_j\, {\rm dv},
$$
where ${\bf B}_i$-vectors are free decay modes (asterisk means complex
conjugation), and integra-tion occurs over all volume of the star. $\hat L$
operator describes magnetic field generators and has the form:
$$
\hat L = R_\omega {\rm rot}\, [V \times \cdot] + R_\alpha {\rm rot}
\, (\alpha \cdot )\, ,
$$
where $V$ is a linear velocity of stellar differential rotation and dot
marks positions where respective free decay mode has to be inserted. One can
find expressions for matrix elements $W_{ij}$ as well as technical details
of calculations in the paper of Sokoloff \& Nefedov (2007). We note here
only that the problem was solved in linear axisymmetric approximation and
all variables were as usually rewritten in dimensionless form and expressed
throughout $R_\alpha$ and $R_\omega$ dimensionless numbers, which
characterize intensity of field generation sources, $\alpha$-effect and
differential rotation respectively:
\begin{equation}
R_\alpha = {\alpha R \over \beta}, \hbox{    } R_\omega = {R^2\over
r}\, { \partial \omega \over \partial r }.
  \label{RaRo}
\end{equation}
Here $\omega = \omega (r,\theta)$ is stellar angular velocity, $r$
and $\theta$ are radial coordinate and latitude respectively, $R$ is 
stellar radius and
$$
\alpha = {\tau \over 3}\, <\! {\bf v}\, {\rm rot}\, {\bf v}\! >,
\quad \beta = {\tau \over 3} <\!v^2\!>,
$$
where $v$ and $\tau$ are velocity and turnover time of a convective vortex
correspondingly, and $<...>$ means averaging over the convective vortices
ensemble. Numbers $R_\alpha$ and $R_\omega$ were estimated basing on typical
values of the parameters involved and thus represent the star as a whole --
see Kuzanyan and Sokoloff (1997) for details.

  For $n$ free decay modes dynamo equations (\ref{dyn}) give $n$
eigenvalues and eigenfunctions. For low $R_\alpha$ and $R_\omega$ these
solutions damp, i.e. the real parts of eigenvalues are negative. Enlarging
$R_\alpha$ and $R_\omega$ one can obtain eigenvalues with $\Re e\, \gamma >0$.
Note that for realistic situations $R_\alpha \approx 1$ (Kuzanyan \&
Sokoloff, 1997), the eigenvalues are determined by so-called dynamo number
$D=R_\alpha R_\omega$ (so-called $\alpha \omega$-dynamo). For the sake of
definiteness we choose $R_\alpha = 0.5$ and vary $R_\omega$. 

  Because of the lack of observational information we adopted
"realistic"\, dependences $V(r) \equiv \omega r = 1 - \exp(-r/R),$
$\alpha(r,\theta) = r/R\, \cos\theta$ to calculated $W_{ij}.$ All
quantitative results presented below belong to this particular choice, but
varying profiles $V(r)$ and $\alpha (r,\theta)$ we found that the results
are robust.

   The result of solution of Eq.(\ref{dyn}) can be presented as a dependence
of real part of several leading eigenvalues on dynamo-number $D,$ that
characterized intensity of dynamo action -- see Fig.1. It follows from the
figure that real part of one eigenvalue becomes positive when $D$ exceeds
the critical value $D_{\rm crit} \sim 5000,$ what means that magnetic field
self-excitation occurs.

  We present in Fig.1 real part of several leading eigenvalues of
(\ref{dyn}) as functions of intensity of dynamo action $D$. If $D$ is larger
then the critical value $D_{\rm crit} \sim 5000,$ real part of one
eigenvalue becomes positive what means that magnetic field self-excitation
occurs. We checked all decay modes and selected modes that can be excluded
without qualitative modification of generation process -- Fig.1 is drawn
after such selection. It also follows from the figure that standard
explanation of work of Parker dynamo is not exactly correct: not two ($T_1$
and $P_1$) but five free decay modes really participate in the process:
three poloidal modes $(P_1$, $P_2$, $P_4$) and two toroidal ones $(T_1$ and
$T_3$). Thus both toroidal and poloidal components of magnetic field are
excited in fully convective stars as in the case of solar dynamo.

  Sokoloff \& Nefedov (2007) have used similar low-mode approach to
investigate dynamo process in stars with thin external convective shell,
i.e. in the case of classical Parker dynamo. They found that $P_1$, $P_2$, 
$T_1$ and $T_2$ free decay modes can describe the process of magnetic
field generation in solar type stars. Therefore the set of free decay modes
which play the main role in magnetic field self-excitation are
different in fully convective and in solar type stars.

 From practical viewpoint it is important that the first set includes
$T_3$-mode whereas the second set includes $T_2$-mode. These modes have 
different spatial structure: $T_3$-mode is located almost at the same
latitudinal range as $T_1$-mode (both are proportional to $\sin 2\theta$
where $\theta$ is a latitude) whereas $T_2$-mode is proportional to 
$\sin 4\theta,$ i.e. more shifted to the equator than $T_1$-mode.

  Fig. 2 shows the difference in temporal evolution of toroidal component of
magnetic field $B_\phi$ in both cases. Such kind of diagrams are thought to
be a link between dynamo theory and observations because in solar case
temporal evolution of latitudinal spot's distribution (so called butterfly
diagram) reflects toroidal magnetic field behavior -- see e.g. Obridko et
al., 2006. By the way due to this reason $B_\phi(\theta,t)$-diagrams are
also referred as butterfly diagrams. While $B_\phi$ depends on $r$ as well
as on $\theta$ it appeared that butterfly diagrams for various $r/R$-ratios
are similar to some extent so to fix the idea we plot in Fig.2
$B_\phi(\theta,t)$ function for $r/R=0.9.$ Level lines in the figure are
shown for several values of $B_\phi$ normalized to the maximal value of this
quantity -- see the figure caption.

  As can be seen from Fig.2 our model predicts that the dynamo excites
traveling (in latitudinal direction) wave in solar type star but standing
wave in a fully convective star in the sense that average latitude of
toroidal magnetic field distribution is expected to be almost constant
during the cycle of activity. Toroidal field strength $B_\phi$ varies with
time such as activity becomes negligible at some phases of the cycle. In
contrast, the solar "active latitude"\, migrates from the middle latitudes to
the equator in course of the cycle. Because of the difference in spatial
configurations of the modes $T_2$ and $T_3$ spatial distribution of magnetic
field is shifted polarwards in fully convective star if compare with stars
with convective shell.

  The difference in dynamo action in fully convective stars and thin shell
stars is connected in the particular with different structure of free decay
mode spectrum. For example the spectrum for classical Parker dynamo consists
of singlets only whereas that of fully convective stars is a sequence of
singlets and doublets. It seems naturally to expect that dynamo action in
stars with small radiative core have to be similar to dynamo in fully
convective stars. It would be interesting to find the size of radiative
core, starting from which the dynamo becomes similar to that of stars with
thin convective shell but this problem is out of the scope of the paper.

  The time in Fig.2 is measured in units of diffusion time $\tau =
R^2/\beta,$ where $R$ is stellar radius and $\beta$ is convective
diffusivity. Our calculations give for fully convective star the cycle
length $0.14\tau$ and $0.4\tau$ for Parker dynamo. If to identify the last
value with 11 year solar cycle then duration of activity cycle in fully
convective stars is:
\begin{equation}
T \simeq 4\, {\left( {R     \over R_\odot     } \right)}^2 {\left(
{\beta \over \beta_\odot } \right)}^{-1} \mbox{  ÌÅÔ}.
  \label{T}
\end{equation}

  The dynamo excites magnetic field if generation sources are
strong enough, what means that dynamo number $D$ should exceed some
critical value $D_{\rm crit}$. It can be seen from Fig.1 that  
$D_{\rm crit} \sim 5000$ in the case of fully convective star, what is
substantially larger than in the case of stars with thin convective shell:
$D^{\odot}_{\rm crit} \sim 300$ according to Kuzanyan \& Sokoloff (1997).

  Thus we found in the frame of our model that operation of dynamo is
different in fully convective stars and stars with thin external convective
shell. At the moment the model includes a number of simplifications (axial
symmetry, kinematic approach etc.) but it allows to investigate dynamo
action at this stage drawing away unknown details of stellar structure. The
advantage of our approach is the possibility to remove simplifications step
by step as new information becomes available.

\section*{Photometric activity of the young star V410 Tau}

  Consider now what can observations tell us about magnetic activity of
WTTSs. V410 Tau is the most investigated object among this type of stars.
The star is located in Taurus-Aurigae star formation region distance to
which is $\simeq 140$ pc (Elias, 1978). V410 Tau is a triple system but
radiation of V410 Tau A component -- K4 V star, $\lg T_{ef}=3.663,$ $\lg
(L/L_\odot) = 0.333\pm 0.085$ (White \& Ghez, 2001) -- dominates in the
optical band. Angular distances of B and C components of the system from the
main one are $0.07^{\prime \prime}$ and $0.29^{\prime \prime}$ respectively
(White \& Ghez, 2001), what corresponds to linear distances larger than 10
a.u. Due to these reasons we will suppose that optical light curve of V410
Tau refers to A component and one can neglect with the influence of B and C
components on the process of magnetic field generation in A component.

   By means of theoretical tracks and isochrones of Youn Kil Jung \& Kim
(2007) we found that mass $M_*$ of V410 Tau A is in the range 1.1-1.2
M$_\odot$ and its age is 1-2 Myr. According to Youn Kil Jung \& Kim $M=1.1\,
M_\odot$ star is fully convective at age $t=1$ Myr and at $t=2$ Myr it has a
radiative core with mass $\simeq 10$\,\%\, M$_*.$ The mass of radiative core
of $M = 1.2 M_\odot$ star is $\simeq 1$\,\%\, M$_*$ and $\simeq 17$\,\%\,
M$_*$ at $t=1$ Myr and 2 Myr respectively. If to use calculations of
D$^\prime$Antona \& Mazzitelli (1994) then the age of the star remains
almost the same and its total mass as well as the relative mass of radiative
core appears a bit larger. We concluded therefore that V410 Tau is either
fully convective or mass of its convective envelope exceeds 90\,\%\, M$_*.$

  Variability of brightness and emission in lines of V410 Tau were
discovered in 60-th (Metreveli, 1966; Mosidze, 1970). Rydgren \& Vrba (1983)
found that brightness of the star varies periodically and according to
modern estimations the period is $\simeq 1.87197^d \pm 0.00001$ (Stelzer et
al., 2003). It was found from Doppler imaging that existence of cool spots
with temperature 1000-1500 K below $T_{\rm ef}$ is the reason of periodicity
(Joncour et al., 1994b; Strassmeier et al., 1994; Hatzes, 1995; Rice \&
Strassmeier, 1996). These spots are located in the relatively narrow range
of longitudes and were observed both at polar and equatorial regions
occupying up to a third of stellar surface.

  Doppler imaging of the star were carried out a few times from 1990 to 1993
yr and it gives information about temporal evolution of stellar spots Rice
\& Strassmeier, 1996). It appeared that typical lifetime of individual spots
is of order of one year such as polar spots are larger and live longer than
equatorial spots. Rice \& Strassmeier searched for variations of relative
positions of polar and equatorial spots and concluded that latitudinal
differential rotation of V410 Tau A is $\simeq 200$ times less than solar
one. In addition to cool spots there are regions with $T>T_{\rm ef}$ at the
surface of V410 Tau A but they occupy relatively small area and their
contribution to variations of brightness is negligible.

   Variations of the size, shape as well as distribution over surface and
number of spots is the reason of secular variations of V410 Tau brightness.
Stelzer et al. (2003) analyzed all published photoelectric observations of
the star from 1978 yr to 2003 yr and set hypothesis that the star has
activity cycle with period 5.4 yr. They noted that if the hypothesis is
correct then it is the first example of cyclic activity in young stars.

  In order to check this hypothesis we increased the quantity of
observational data and used wavelet analysis to search for activity cycle.
To do so we investigated $\simeq 400$ photo plates of Sternberg Institute's
"glass library"\, obtained during 1905-1987 yr time interval. We digitized
the plates by means of CREO scanner which has 2540 dpi resolution (Samus et
al., 2006). The subsequent data processing was carried out in a standard way
such as we used 20 stars for comparison assuming that color system of
photo plates is close to that of photoelectric observations in $B$-band.
\footnote{
There are both photographic and photoelectric data in the JD
2.442.300-2.446.700 time interval. We concluded from the comparison of these
data -- see Fig.5 -- that color system of photo plates is close enough to
$B$-band filter in order to use photographic and photoelectric data for
joint analysis. 
} 
B-magnitudes of comparison stars were adopted from USNO A2.0 catalog {\it
http://simbad.u-strasbg.fr/Simbad}. We estimated that mean square root error
of our measurements is $\simeq 0.07^m.$

  We added to our photographic measurements B-band photoelectric
observations which were adopted from {\it
http://www.astro.wesleyan.edu/pub/ttauri} database (Herbst et al., 1994).
Summary light curve of V410 Tau is shown in Fig.3 from which one can see
that at large time interval the brightness of the star varies in more
complicated way if compare with the period of photoelectric observations.
The main new feature of the light curve is the large scale decreasing of
brighntness at $\Delta B \simeq 0.4^m$ near 1960 yr. There are no photographs
of the vicinity of V410 Tau in the Institute's archive during 1913-1950 yr
period, so we can state only that the average brightness of the star at the
first decade of XX-th century was nearly the same as during 1965-1985 yr
time interval, i.e. after the period of minimal brightness. It can be seen
from Fig.3 that after 1985 yr the average brightness of the star slightly
increased and the amplitude of seasonal variability began to vary in a wavy
way what was interpreted by Stelzer et al. (2003) as a cycle of activity
with 5.4 yr timescale.

  The existence of similar wavy modulation of the light curve is not obvious
before 1985 yr. To answer the question about existence or absence of cyclic
activity in V410 Tau we carried out wavelet analysis of its light curve using
so called adaptive method described in the paper of Frick et al. (1997a).
One can also find in this paper explanation of advantage of using wavelet
analysis in investigation of stellar activity cycles. Results of the
analysis are presented in Fig.4 where temporal (X-axis, JD) variation of
amplitude of harmonics of different frequency $\nu$ (plotted along Y-axis)
is shown. Relative value of harmonic's amplitude are shown with different
hues of grey color such as the darker color the larger the amplitude. For
convenience frequency axis is labeled in units of harmonic's period
$T=1/\nu$ expressed in years.

  It appeared that the four regions (three strips and the spot) have the
largest intensity -- they are marked with white lines and numerated from 1
to 4 in Fig.4. The spot 3 obviously reflects the existence of the 1960 yr
minimum on the light curve and strips 1,2 \& 4 can be considered as quasi
harmonics with non-stable period. Strip 1 corresponds to a set of harmonics
with periods 4-5 years and respects to the temporal interval within which
cyclic activity with 5.4 yr period was observed as Stelzer et al. (2003)
stated. It can be seen from the figure that in fact it is more correct
to speak about quasi cyclic activity. 

  Harmonics with periods 8-9 yr (Strip 2) and 11-13 yr (Strip 3) are
strong within 1950-1990 time interval. Amplitude of harmonics of Strip 3 is
the largest what is in agreement with Fig.3: the low envelope of the
light curve just has local minima that repeated with the same time interval.
But total number of cycles with such timescale is not large enough during
the period of observations to be sure that they are really exist. To check
if V410 Tau indeed has the cycle with quasi period 11-13 yrs it is necessary
to use data from "glass libraries"\, of other observatories.

  Let consider temporal variations of V410 Tau's phase light curves converged
with rotation period as an additional source of information about secular
variations of activity of the star. It can be seen from Fig.5 that
variations of the phase light curve's shape were relatively small from 1987
yr to 2005 yr: at that period the phase light curve looked as one hump curve
with maximum near 0.5 phase -- see the bottom panel in the right column.
Judging from results of Doppler imaging obtained in the first half of 90-th
such form of the light curve was the result of existence of "active longitude
strip"\, inside of which large size cool spots were concentrated.

  It is reasonable to assume that active longitude strip also existed at
those observational seasons when phase light curve had the form of one hump
curve. At the first time such form of the curve were observed at the global
minimum epoch i.e. from 1958 yr to 1962 yr, then at 1972-1976 observational
seasons i.e. 10-11 years later and then disappeared again to appear once more
9-10 years later. From that moment, i.e. approximately from 1985 yr and up
to 2005 yr the phase curve almost saved its form but moved quasi cyclically
as a whole along the phase relative average position with timescale of order
of 4-5 years. \footnote{ It can be seen from Fig.3 and 4 of Stelzer et al.
(2003) and also explains diffuse form of the phase curve in our Fig.5 at
that epoch.} Note that one hump phase curve also existed during $\simeq 4$
years before 1985 yr epoch, so one can assume that it is typical lifetime of
magnetic field configurations which produce active longitude strip.

  An important information concerning surface distribution of spots on the
surface of V410 Tau can be extracted from photometric observations even for
epochs when the form of phase light curve differed from one hump curve while
no Doppler imaging of the star exist before 90-th. In the particular
brightness of V410 Tau varied slightly and chaotic around average value
$B\simeq 12.2-12.3$ at 1963-1970 yr, i.e. the star was then a bit more
bright than at minimum of brightness at 90-th yr epoch -- see Fig.5. It
means that small amplitude of variability at 60-th was due to relatively
homogeneous spot's distribution over longitude rather than the absence of
spots. At 1975-1984 yrs epoch the amplitude of variations was nearly the
same but the phase light curve looked more regular. For example the phase
light curve had two humps at 1981-1984 yrs epoch and Herbst (1989) could
reproduce this shape assuming that there are two large spots at V410 Tau's
surface situated at different hemispheres and shifted at $\simeq 180^o$ over
longitude relative each other.

 Thus the following picture of spot's dynamics during the last fifty years
can be drawn. A powerful group of spots has appeared at the surface of the
star at the end of 50-th yrs of XX century such as it was situated within
some strip of ("active") longitudes. One can not conclude from the existing
data what was the reason of significant decreasing of stellar brightness at
that epoch: increasing of total surface occupied by spots or decreasing of
their average temperature or both factors. In any event the group
disappeared after $\simeq 4$ years of existence and spots redistributed more
or less homogeneously over stellar surface.

  Due to some reason longitudinal spot's distribution suddenly (during
$\Delta t < 1$ yr) became strongly non-homogeneous again $\simeq 10$ years
later such as the position of "activity stripe"\, shifted at $\simeq 120^o$
over longitude relative previous position. Clearly expressed non-homogeneity
of longitudinal spot's distribution existed near 4 years and then it
disappeared during a few monthes. Longitudinal distribution of spots was
almost homoge-neous during subsequent 3-4 years but then slow process of
spot's grouping over longitudes has began. As a result the next strip of
active longitudes has formed after 4-5 years (i.e. at the middle of 80-th
yr) which in contrast to previous ones was stable enough: its evolution
during subsequent 20 years is reduced to small ($\sim 0.1$ along phase)
regular vibration over longitude relative some average position.

  We concluded from all written above that secular variations of V410 Tau's
brightness is the result of some processes that produce redistribution of
cool spots over longitudes. It will be possible to decide if this conclusion
is correct to all WTTSs only after analysis of large enough number of light
curves of other stars. It is non-trivial task because judging from V410 Tau
case one needs observational set of 50-100 yr duration what means that
photographic estimations of brightness should be included in the analysis,
but accuracy of such data $(\sim 0.1^m)$ is comparable with amplitudes of
variability of most WTTSs (Grankin et al., 2008).

\section*{The theory and observations: what and how to compare?}

  As we noted dynamo excites magnetic field if dynamo number $D=R_\alpha
R_\omega$ exceeds some critical value $D_{\rm crit},$ which is according to
our estimations $\sim 5000$ in the case of fully convective stars. For the
Sun $D \sim 10^3-10^5$ (see e.g. Kuzanyan \& Sokoloff, 1997). According to
Eq. (\ref{RaRo}) $D$ is proportional to the product of angular velocity to
its radial gradient. The first term is 10 times larger for WTTSs than in the
Sun however the second term is unknown. We mentioned in the
Introduction that the differential rotation $\partial \omega /\partial
\theta$ for young stars is thought to be much weaker than solar, but
it does not mean that the {\it radial gradient} of angular velocity is also
weak. Because $\partial \omega /\partial r$ is unknown we can not insist
that the magnetic field of WTTS stars have to be excited by dynamo. Thus
estimation of dynamo number for young stars based on observational data
and/or theoretical models is very important problem.

  Even if the concept of dynamo origin for magnetic field of young stars
would be firmly established, comparison between the dynamo theory and
observational data remains a complicated problem. The point is that the
axisymmetric dynamo models (numerical calculations or our low-mode
approximation) predict a cyclic activity in the form of periodic variations
of {\it latitudinal} distribution of stellar spots (see Fig. 2). In contrast
observed secular variations of WTTS V410 Tau's light curve seems to be
associated with (quasi periodic) {\it longitudinal} redistribution of
starspots. Similar processes of longitudinal evolution of spots distribution
is also known in solar case (Berdyugina et al., 2006) but they are much less
pronounced than that of V410 Tau. Longitudinal evolution of solar spots
distribution is connected with fine features of solar dynamo action and
therefore in the case of nearly axisymmetric situation one have to compare
predictions of axisymmetric dynamo models with evolution of latitudinal
magnetic field (or stellar spots) distribution.

  One can not extract information about latitudinal (re)distribution of
starspots during the process of arising and disappearing of active
longitudes from photometric observations only -- Doppler imaging of WTTSs
during decades are necessary for this aim. In other words it is impossible
at this stage to get V410 Tau's butterfly diagram of substantial duration
from observation and confront it with theoretical predictions. Unfortunately
such contradiction between abilities of observations and theoretical demands
is a general problem in stellar cyclic activity studies.

  It means in the particular that there is no sense at this stage to compare
our theoretical estimation (\ref{T}) of activity cycle duration with
timescales of V410 Tau's light curve secular variations. The comparison will
become meaningful after a clarification of the role of effects connected
with inclination of the magnetic axis to the rotation one. As we noted
earlier it is used to suppose that magnetic and rotational axes of T Tauri
stars make significant angle (Bouvier et al., 2007), however Donati et al.
(2007, 2008) found recently that in some stars this angle can be $<10^o.$
Comparison theoretical predictions with observations is more reliable for
such stars, but up to date we did not collect enough photometric data to
analyze their long term variability.

  Our theoretical model (low mode approach) while it is relatively simple
predicts qualitative difference between dynamo cycles in fully convective
stars and in stars with thin convective shell. At first it follows from our
calculations that latitudinal distributions of spots during activity cycle
are different in stars of these two classes. The second, the model predicts
that spot generation rate have to be very low at some phases of activity
cycle in fully convective stars. But from quantitative viewpoint these
results should be considered with care at least because it is unknown to
what extent adopted profiles $V(r)$ and $\alpha (r,\theta)$ are realistic.

  The lack of this information does not allow to explain one of the main
peculiarity of WTTS stars: extremely high level of their activity.  One can
suppose that magnetic field generated in fully convective stars is
concentrated in more external layers if compare with solar type stars and it
is why it reaches stellar surface easier. More information about
differential rotation of young stars is necessary to confirm or reject this
hypothesis.

   One should not forget also about simplified approximations in the frame
of which we used low mode approach to solve the problem of magnetic field
generation in fully convective stars. Let us remind that the problem was
solved in axisymmetric and kinematic approximation. The last means that no
back reaction of magnetic field on the gas motion in convective zone was
taken into account, what looks unreasonable bear in mind that observed field
strength in T Tauri stars is $\sim 3$ kG. The effects of back reaction
hardly can be predicted in advance. E.g. Tworkowski et al. (1998) found for
solar type stars that effects of back reaction in convective zone results in
substantial variations of the amplitude from one cycle to the other. One can
not also exclude the appearance of activity in which chaotic component will
exist along with cyclic one.

\section*{Conclusion}

  Calculations of magnetic field generation process which we carried out in
the frame of low mode approach as well as numerical calculations (e.g.
Dobler et al., 2006) indicate that dynamo can work in fully convective stars
creating periodically varying magnetic field. Assuming "realistic"\, low of
stellar differential rotation we found that dynamo starts to work in fully
convective stars when dynamo number $D>5\cdot 10^3$ and estimated the period
of the cycle as well.

   Toroidal component $B_\phi$ of magnetic field is excited in fully
convective stars along with poloidal component as in the case of stars with
thin external convective shell. We tried to compare our theory with
observations of young star V410 Tau (WTTS type) assuming that $B_\phi
(\theta,t)$ dependence corresponds to time evolution of latitudinal starspots
distribution. Most WTTSs are fully convective and their activity level
highly exceeds the solar one and differs from the later in some respects.

  It follows from our calculations that dynamo generated cycle in fully
convective stars should differ significantly if compare with thin convective
shell stars: 1) the latitudinal distribution of star spots during the cycle
in both cases is expected to be different; 2) the model predicts that the
spot formation rate in fully convective stars becomes very weak at some
phases of the cycle. The above difference seems to be due to the difference
in the free decay mode spectra in both cases. It would be instructive to
clarify how the difference under discussion depends on the size of the
radiative core.

   It follows from our analysis of V410 Tau light curve that the character
of photometric activity of the star during $\sim 50$ yr time interval is
much more complicated than it was found from investigation of solely
photoelectric observations which cover time interval less than 30 years.
It appeared that photometric activity of the star is not a definitely
expressed cycle with fixed period similar to solar one. Rather it can be
described as a quasi periodic process with timescale $\sim 4$ yr. with
chaotic component over imposed. It is unclear if V410 Tau has
(quasi) periodic component with timescale of order of 11 yr.

  We concluded from the analysis of phase light curves at different epochs
that secular variations of stellar brightness is the result of
redistribution of star spots over longitude. This circumstance does not give
possibility to compare observations with predictions of our model, which was
developed (for simplicity) under assumption of axial (longitudinal) symmetry
to investigated temporal evolution of latitudinal spot's distribution, but
one can not extract just this information from photometric observations.

  It is unknown to what extent activity of V410 Tau is typical to T Tauri
stars without accretion discs. To clarify the problem it is necessary to
carry out analysis of light curves of many young stars and to organize the
program of systematic Doppler imaging of a few WTTSs over some decades.

  In order to understand how large the difference between two kinds of
dynamo under consideration and if this difference can indeed explain
observed peculiarities of WTTS's activity it is necessary to consider dynamo
models which take into account non-linear effects, deviation from axial
symmetry and of course based on realistic lows of differential rotation of
investigated stars -- we plan to do all this in the future. The advantage of
our approach is possibility of its straightforward generalization by means of
incorporation of all these improvements step by step in a relatively simple
way. Subsequent comparison of predictions of dynamo theory with observations
looks very promising.

 {\it Acknowlegements.} The work was supported by RFBR grant 04-02-16068 and
INTAS grant 03-51-6311. We thank our colleagues from Variable star devision
of Sternberg astronomical institute for the help with scanning of plates of
the institute's "glass library", Drs. K.Grankin and W.Herbst who provided us
with photoelectric observational data and Drs. Youn Kil Jung and Y.-C. Kim
for information concerning parameters of young star's models.


\newpage

{\centerline {\Large {\bf Figures }}}

\begin{figure}[h!]
 \begin{center}
  \resizebox{14cm}{!}{\includegraphics{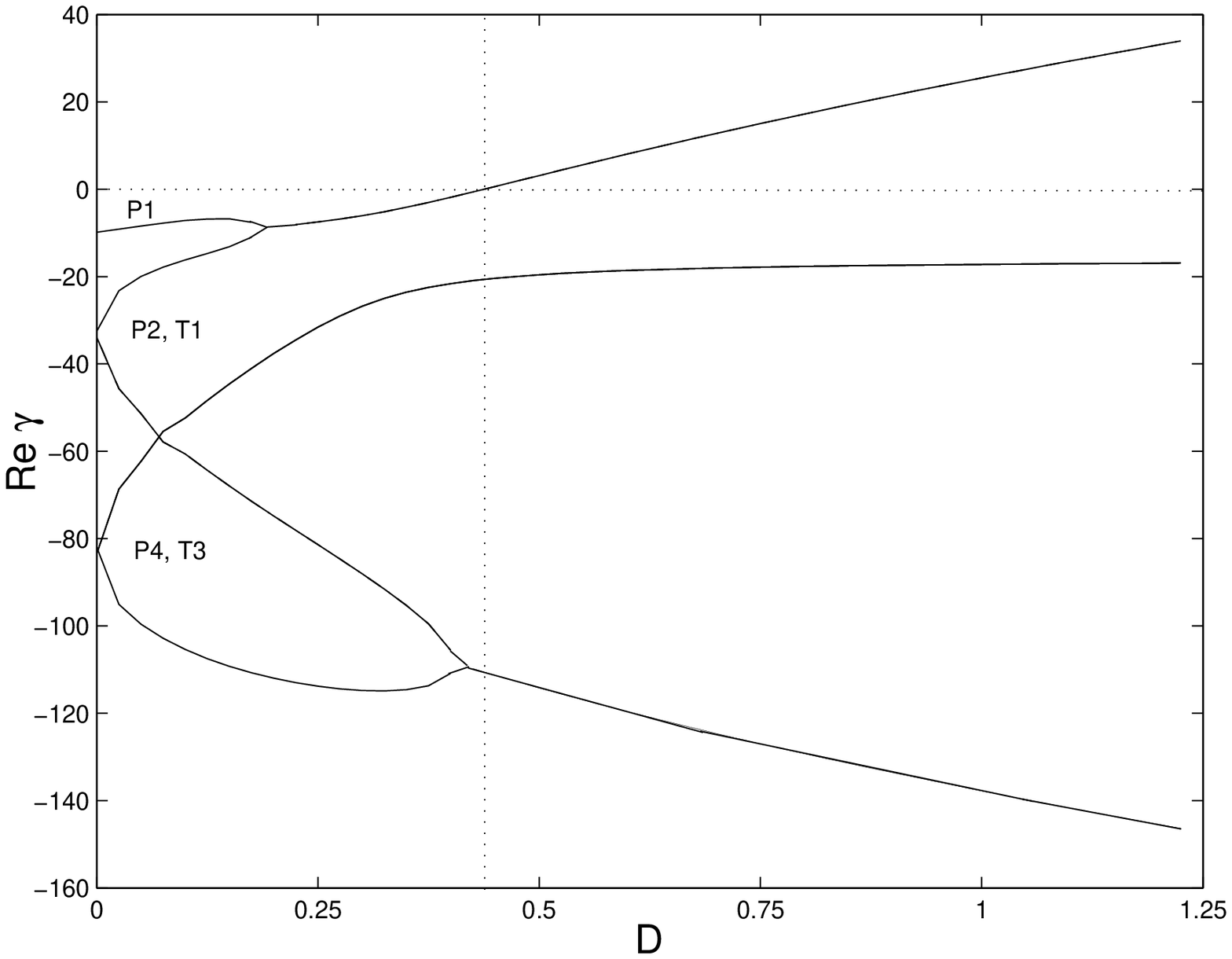}}
  \caption{
The dependence of real part of the leading eigenvalues on dynamo-number $D,$
expressed in $10^4$ units. Labels near curves indicate which free decay
modes give the main contribution to respective eigenfunction. Vertical
dotted line marks the value of $D$ starting from which magnetic field self
generation occurs.
          }
 \end{center}
\end{figure}

\begin{figure}[h!]
 \begin{center}
  \resizebox{12cm}{!}{\includegraphics{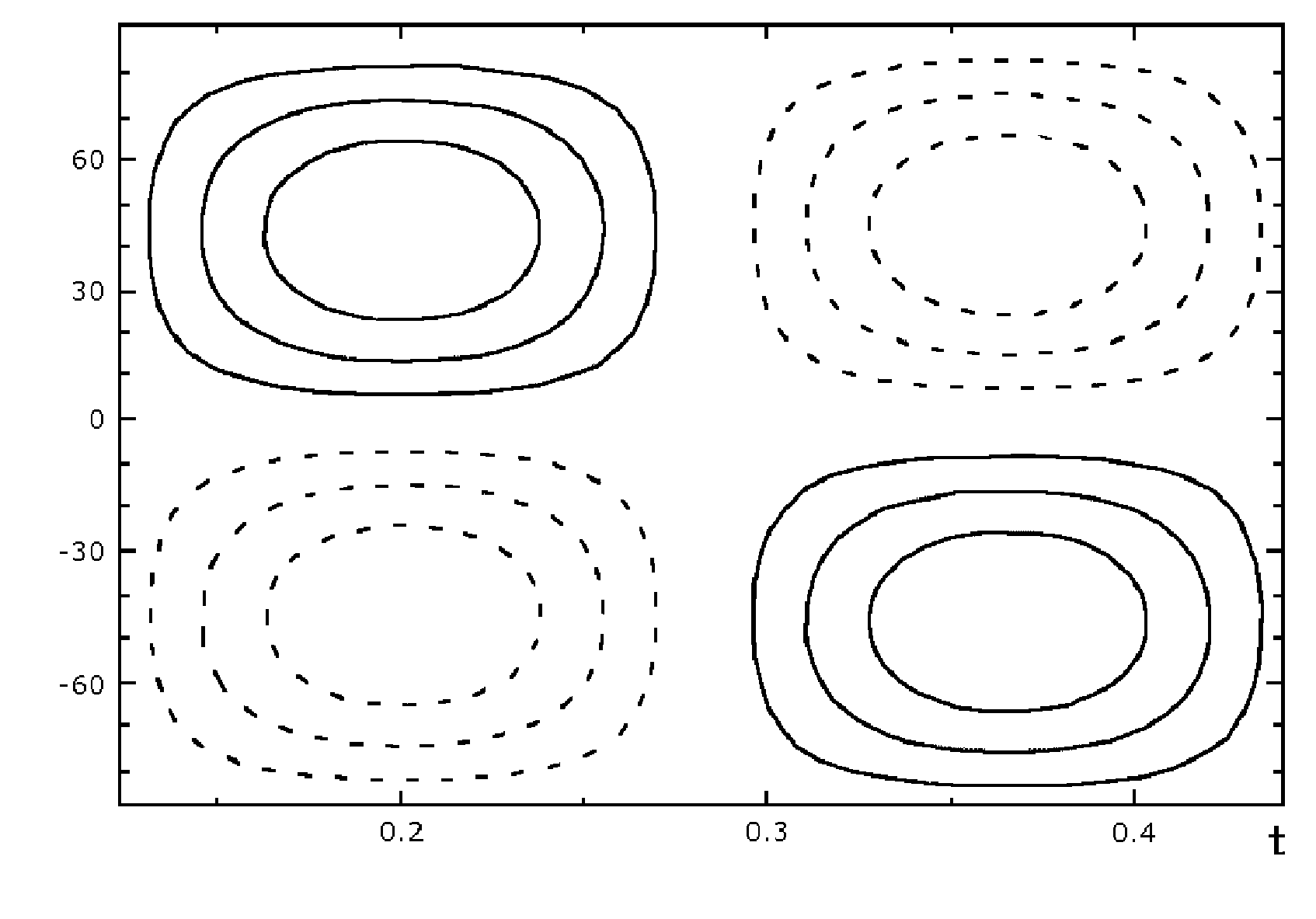}}
  \resizebox{12cm}{!}{\includegraphics{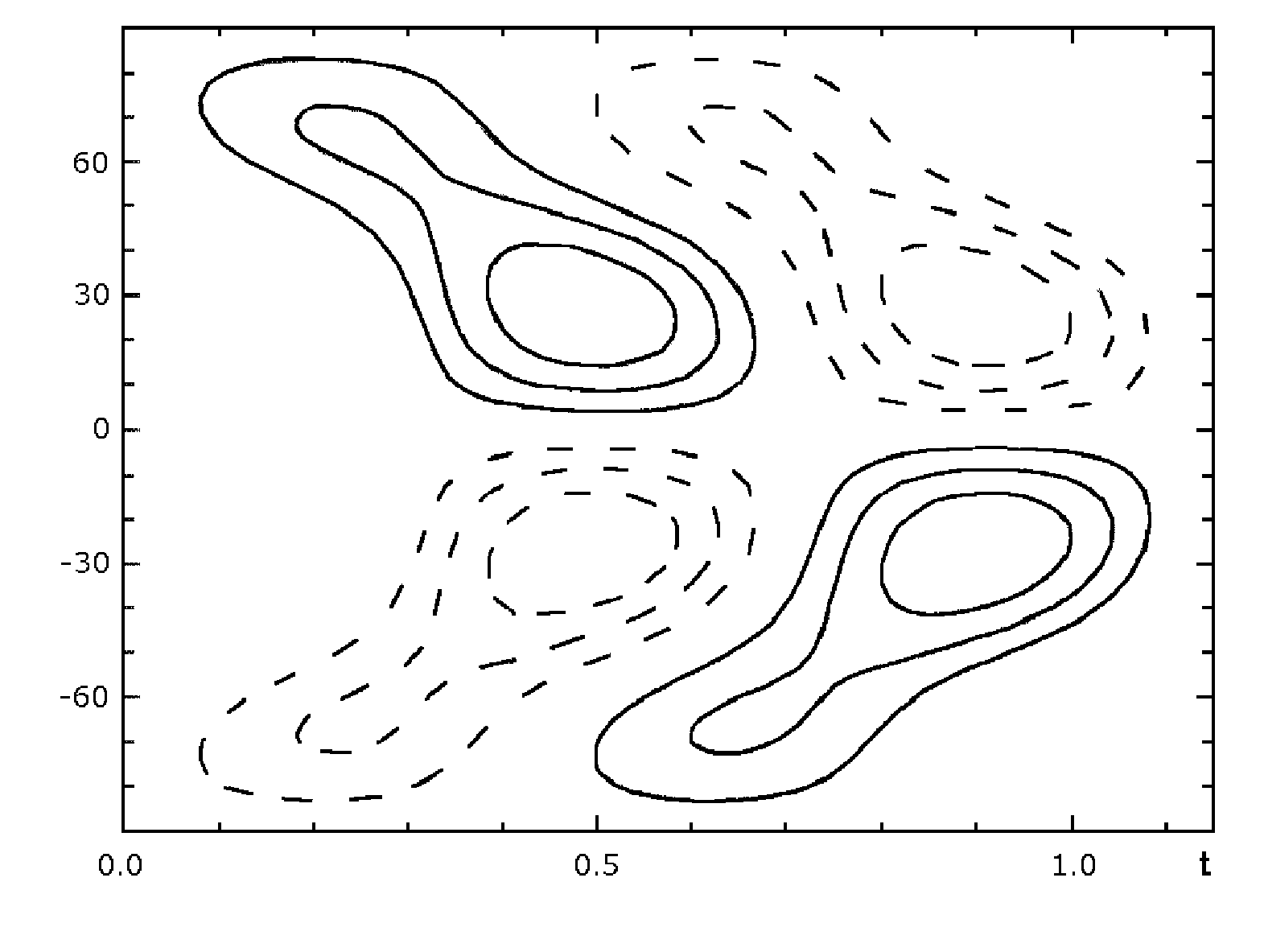}}
  \caption{
Variation of magnetic field's toroidal component $B_\phi$ with time (X-axis)
at different latitudes (Y-axis) for fully convective stars (top panel) and
for stars with thin convective shell (bottom panel). Time and latitude are
expressed in units of diffusion time $\tau$ and degrees respectively. Level
lines for $B_\phi > 0$ are shown with thick lines and for $B_\phi<0$ with
dashed lines. The lines are drawn for the following values of $B_\phi$
normalized to maximal value $B_\phi^{max}$ (from the center to the
pereferee): $\pm 0.7,$ $\pm 0.4,$ $\pm 0.15$ at upper panel and $\pm 0.8,$
$\pm 0.5,$ $\pm 0.15$ at lower panel. See text for details.
           }
 \end{center}
\end{figure}

\begin{figure}[h!]
 \begin{center}
  \resizebox{16cm}{!}{\includegraphics{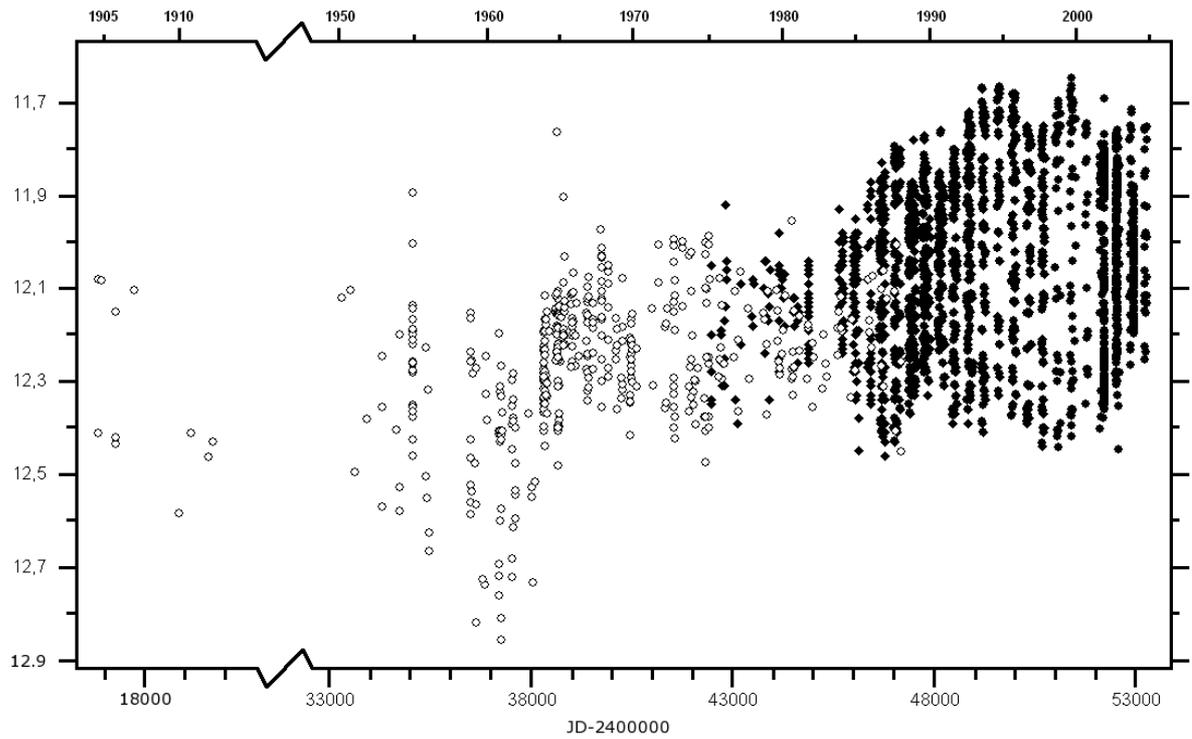}}
  \caption{
B-band light curve of V410 Tau based on photographic (empty dots) and
photoelectric (filled dots) observations. See text for details.
           }
 \end{center}
\end{figure}

\begin{figure}[h!]
 \begin{center}
  \resizebox{16cm}{!}{\includegraphics{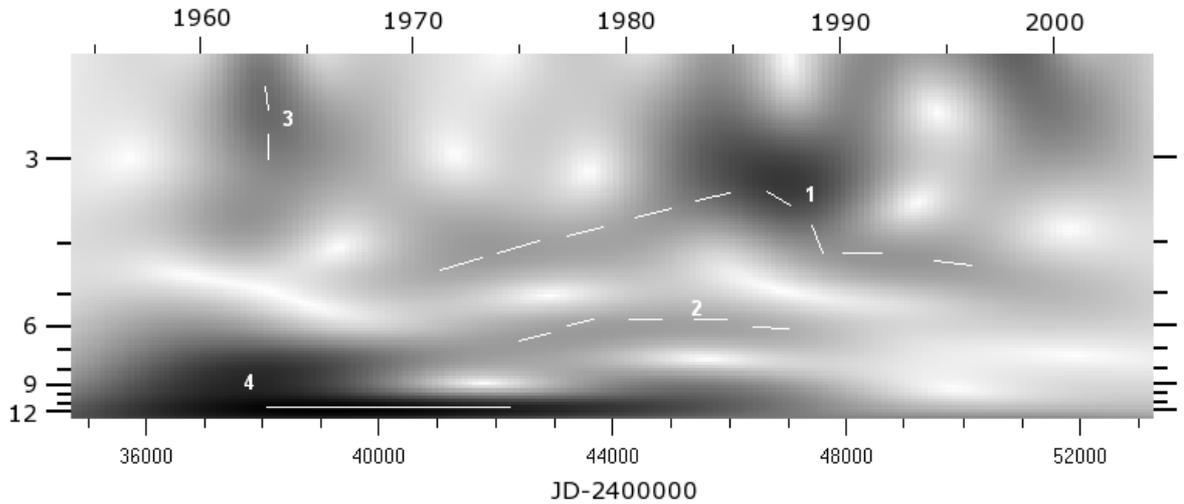}}
  \caption{
Results of wavelet analysis of V410 Tau's light curve. Time and frequency of
wavelet-harmonics $\nu$ are plotted along X- and Y-axes respectively, but
vertical axis is labeled in periods $T=1/\nu$ (expressed in years) for
convinience. See text for details.
           }
 \end{center}
\end{figure}

\begin{figure}[h!]
 \begin{center}
  \resizebox{16cm}{!}{\includegraphics{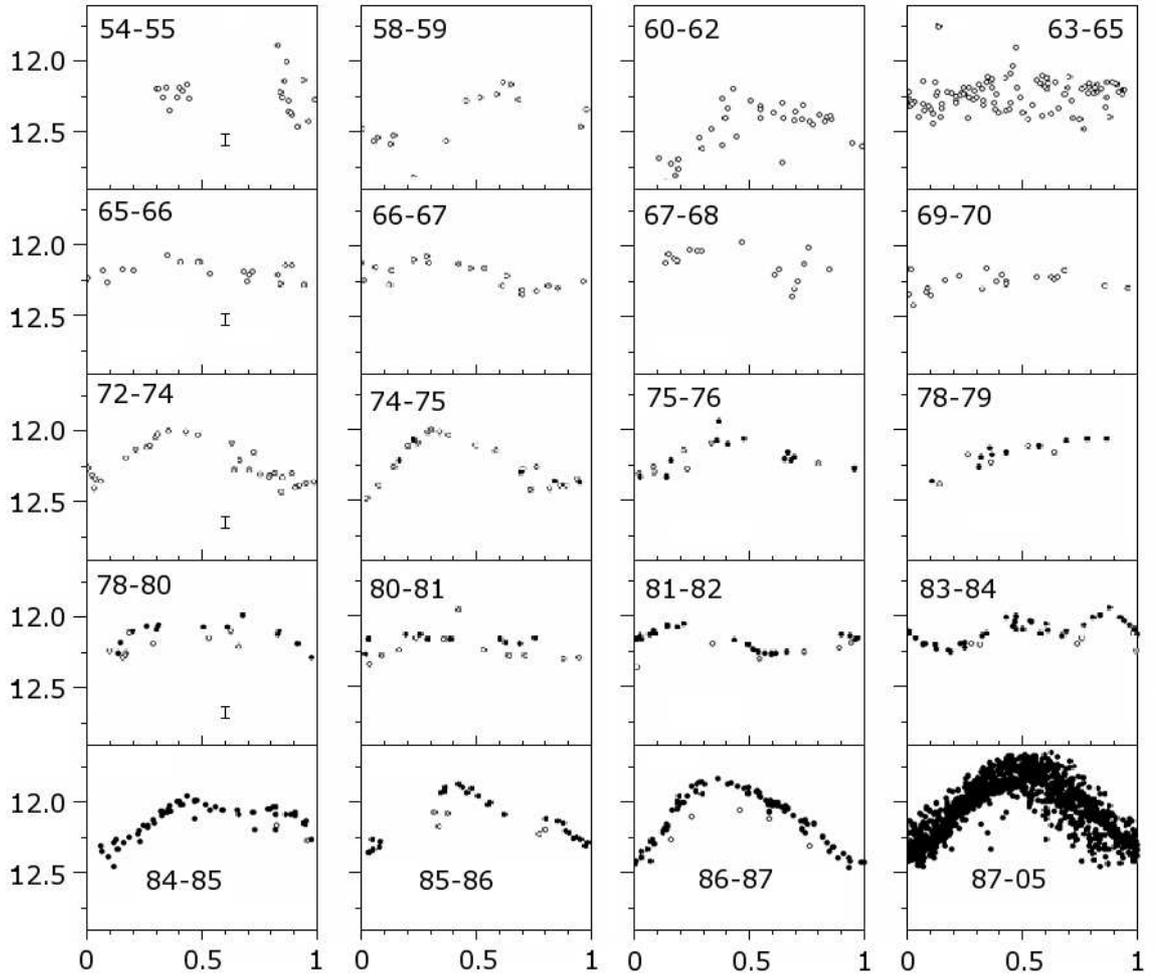}}
  \caption{
Phase light curves of V410 ôau at different observational seasons (written at
each panel). Phase and B-magnitude are plotted at horizontal and vertical
axes respectively. Error bar shown at panels of the left column indicates
error of photographic measurements (empty dots), and the error of
photoelectric observations (filled dots) is $\simeq 3$ times less. See text
for details.
           }
 \end{center}
\end{figure}

\end{document}